\documentclass[10pt,conference]{IEEEtran}
\IEEEoverridecommandlockouts
\usepackage{cite}
\usepackage{amsmath,amssymb,amsfonts}
\usepackage{algorithmic}
\usepackage{fancyhdr}
\usepackage{graphicx}
\usepackage{textcomp}
\usepackage[table]{xcolor}
\usepackage{pdfpages}
\usepackage{bm}
\def\BibTeX{{\rm B\kern-.05em{\sc i\kern-.025em b}\kern-.08em
    T\kern-.1667em\lower.7ex\hbox{E}\kern-.125emX}}
    
\fancypagestyle{specialfooter}{%
\fancyhf{}

\fancyfoot[R]{ \noindent\fbox{%
\parbox{\textwidth}{%
{\footnotesize \bf \tiny{© 2024 IEEE.  Personal use of this material is permitted.  Permission from IEEE must be obtained for all other uses, in any current or future media, including reprinting/republishing this material for advertising or promotional purposes, creating new collective works, for resale or redistribution to servers or lists, or reuse of any copyrighted component of this work in other works.}}
}
}}
}

\begin{document}


\twocolumn

\title{Fast variational   knowledge graph embedding}

\author{\IEEEauthorblockN{Pulak Ranjan Giri}
\IEEEauthorblockA{
\textit{KDDI Research, Inc.}\\
Fujimino-shi, Saitama, Japan \\
pu-giri@kddi-research.jp}
\and
\IEEEauthorblockN{Mori Kurokawa}
\IEEEauthorblockA{
\textit{KDDI Research, Inc.}\\
Fujimino-shi, Saitama, Japan \\
mo-kurokawa@kddi-research.jp}
\and
\IEEEauthorblockN{Kazuhiro Saito}
\IEEEauthorblockA{
\textit{KDDI Research, Inc.}\\
Fujimino-shi, Saitama, Japan \\
ku-saitou@kddi-research.jp}
}

\maketitle\thispagestyle{specialfooter}

\maketitle

\begin{abstract}
Embedding  of a  knowledge graph(KG)  entities and relations   in the form of  vectors  is an  important aspect for  the  manipulation of  the  KG database for several downstream tasks,  such as  link prediction, knowledge graph completion, and  recommendation.  Because of the growing size of the knowledge graph databases, it has become a daunting task for the classical  computer to train a model efficiently.   Quantum computer can  help  speedup  the embedding process of the KGs  by  encoding the entities  into a variational quantum circuit of polynomial depth. Usually,  the  time complexity for   such  variational circuit-dependent quantum classical algorithms  for each epoch   is  $\mathcal{O}(N \mbox{poly}(\log M))$, where $N$ is  number of elements in the knowledge graph and $M$ is the  number of features  of each entities  of the  knowledge graph.  In this article we exploit  additional  quantum advantage by training multiple elements of KG  in superpositions, thereby reducing the computing time  further for the knowledge graph embedding model.

\end{abstract}


\section{Introduction} \label{in}
In the present era of advanced  computing,  knowledge graph(KG) database  has been a good source of information, which can be exploited  for the benefit of the society. However, we need to embed the the elements of the KG database so that it can be manipulated for several useful  tasks,  such as,   link predictions, subgraph classifications,  and question answering in artificial intelligence(AI).   Although, classical knowledge graph embedding(KGE)  models \cite{wang}, such as  RESCAL, TransE,  DistMult, TuckER, ComplEx, ConvE and NeuralLP, work well, but their  will eventually take extremely long time  train   for  very  large  KG database, making them inefficient. 

One of the suitable choices  to overcome the limitations related to  handling large  KGs, is to exploit quantum computing technique for embedding the KG elements to the associated Hilbert space.   For example,  a quantum counterpart of the RESCAL model \cite{nickel},  studied in ref. \cite{ma},  uses variational quantum circuit  to embed head, tail and relation of a KG database.  Expressibility and entangling capacity of the quantum circuit \cite{mori} used in this model  and   quality and quantity of  negative triples \cite{giri}  play a crucial role in its performance.  Training time for  such variational quantum circuit is  polynomial in  $\log M$ and linear in $N$, where $M$ and $N$ are  the number of features and number of elements  of a  KG database.  Even most of the hybrid quantum classical models have time complexity linear in $N$. 

In this article, we further reduce the time complexity of the quantum knowledge graph embedding model by simultaneously using  multiple triples at a time  to train the model.  
We   arrange  this article  as follows:  A brief discussion  on  quantum knowledge graph embedding with variational circuit   is provided   in section    \ref{qkge}.   In  sections    \ref{qram}  and \ref{expr}  fast knowledge graph embedding with quantum variational circuit   and the corresponding  experimental results are  discussed  respectively.    Finally  we conclude with a discussion in section \ref{con}.

\section{Quantum knowledge graph embedding} \label{qkge}
A KG element, known as fact/triple$(h,r,t)$, is generally  composed of  head$(h)$, relation$(r)$ and tail$(t)$.  For example, consider the triple ``Tokyo is the capital of Japan". Here  ``Tokyo" is the head, ``is the capital of" is the relation and ``Japan" is  the tail.   In KGE, entities(head,tail) and relations are encoded as the  vectors of a low dimensional vector space  and matrices acting on the vectors respectively.   Depending on the specific KGE model both entities and relations can be   encoded as vectors as well.   In RESCAL model  $h$ and $t$ are encoded as the $d$-dimensional vectors and $r$ is encoded as a $d\times d$ matrix, $M_r$. The model is trained by choosing the scoring function $h^T M_r t$. In TransE model however,   $h$, $r$ and $t$ are encoded as the vectors of a vector space.   The model is then  trained by choosing the scoring function $|| h + r - t ||$, where  $|| . ||$ is the  $L_1$ or $L_2$ norm.

In the quantum model for knowledge graph embedding,  entities are  embedded as the quantum states on a Hilbert space, $\mathcal{H}$ and relations are embedded as the operators acting on the Hilbert space.  Let us assume that  there are  $N_e$ entities,  $ h, t \in \{ e_1, e_2, \cdots, e_{N_e} \}$,  and  $N_r$ relations  $ r \in \{ r_1, r_2, \cdots, r_{N_r} \}$. 
The entities   are   represented by   the $n$-qubit quantum state   $|e_i \rangle$  as 
\begin{eqnarray}
 |e_i \rangle = \mathcal{U}(\bm{\alpha}_{e_i})H^{\otimes n} |0 \rangle^{\otimes n}\,, 
\label{head}
\end{eqnarray}
where  $\mathcal{U}(\bm{\alpha}_{e_i})$ is  the variational  quantum circuit for the  entity, $e_i$,  with parameter    $\bm{\alpha}_{e_i} = \left[\alpha_{e_i}^1, \alpha_{e_2}^2, \cdots, \alpha_{e_i}^{n_e}\right]$.    
Variational circuit  for the KGE model in ref. \cite{ma}  is trained by distinguishing  positive triples from the negative triples \cite{giri}. 
The scoring function  for a triple $(h, r, t)$ 
\begin{eqnarray}
\delta_{hrt} =  |\langle t | \mathcal{U}(\bm{\beta}_{r}) |h \rangle|^2 \,,
\label{sc}
\end{eqnarray}
can be used to train the model.  Unitary operator  $\mathcal{U}(\bm{\beta}_{r_i})$ in eq. (2), with  the  parameter  $\bm{\beta}_{r_i} = \left[\beta^1_{r_i}, \beta^2_{r_i}, \cdots, \beta^{n_r}_{r_i}\right]$ represents  the relation $r_i$. Usually, quantum circuits with high expressibility and entangling capacity are suitable to represent  $\mathcal{U}$s.
Aim of the KGE model  is to train it  in such a way that   the head  align in a way  so that scoring  reaches to  $\delta_{h r t} \to 1$,  for the positive  triples and  $\delta_{h r t} \to 0$,  for the negative triples.  
A classical optimisation  technique is then used to train the model  by  minimizing   a suitably chosen  loss function, such as  mean squared  error   loss function: 
\begin{eqnarray}
L = \frac{1}{D} \sum_{h,r,t}\left(  \delta_{hrt} - y_{hrt} \right)^2 \,,
\label{loss}
\end{eqnarray}
where $D$ is  number of triples  in a batch for training and $y_{hrt}$ is the  label  corresponding to the triple $(h,r,t)$.

\section{Fast QKGE} \label{qram}
Usually quantum machine learning models have complexity advantage with respect with the  features, $M$,  of a database. However, it  is   possible to train multiple samples  in a superposition as in ref.  \cite{dan}, which  can  have a further complexity advantage with respect to the number of samples, $N$.  If we can train all the $N$ samples simultaneously for $m$ epochs using a QRAM, the complexity for the model with respect to $N$ becomes  $\mathcal{O}(m\log N)$, which is faster than the standard complexity 
of $\mathcal{O}(mN)$.  Total time complexity then becomes  $\mathcal{O}(m\log N \mbox{poly}(\log M)/\epsilon^2)$, where $\epsilon$ is the accuracy of the model. 

In the article, multiple triples are simultaneously used for training  the knowledge graph embedding model. Specifically, we use FFQRAM \cite{park} to  encode  the KG entities. In fig 1, a    variational  quantum circuit to simultaneously use  four triples  is presented, where  Pennylane's  inbuilt-template of  strongly-entangling-layer  is  used to represent  entities on a $4$-qubit  quantum state and another  strongly-entangling-layer   is used to represent relation. 
Parameters of both entities and relation are optimized  during training and  the performance of the trained model is evaluated. 


\begin{figure}[h!]
  \centering
     \includegraphics[width=0.40\textwidth]{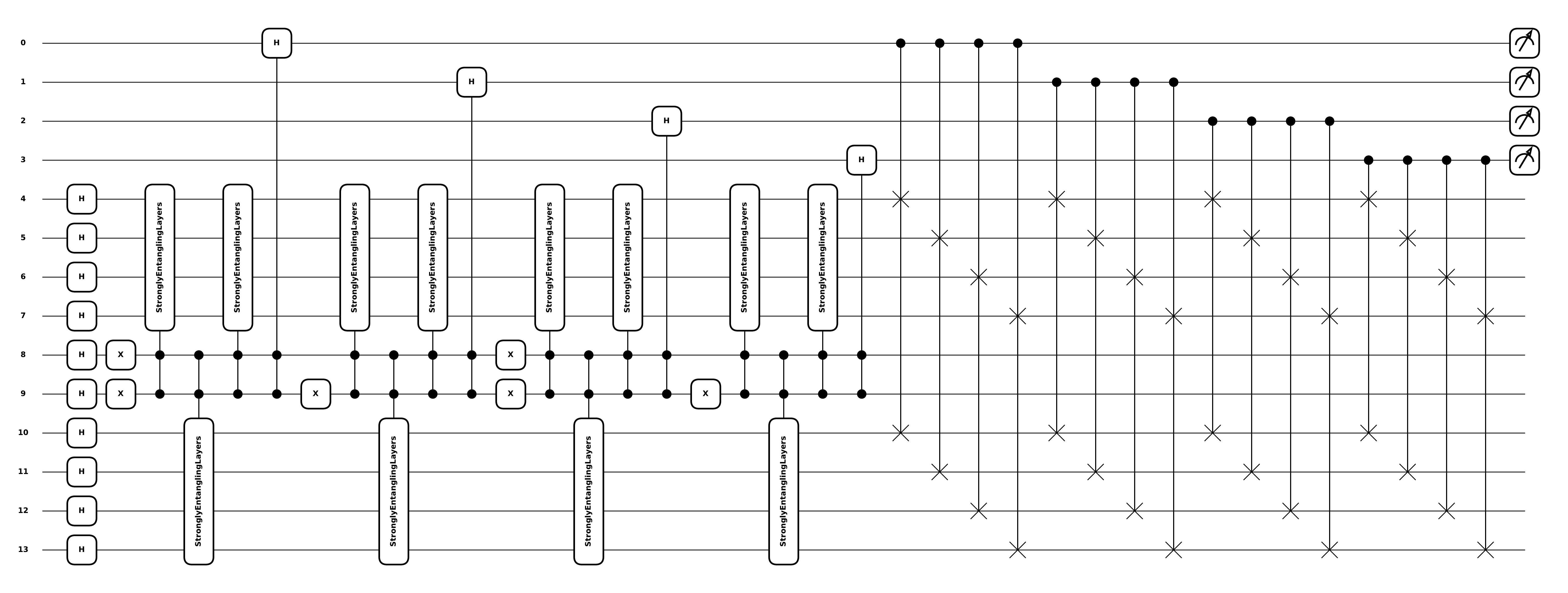}
          
       \caption{Variational quantum  circuit for  knowledge graph embedding.}
\end{figure}
\section{Experimental results} \label{expr}
\subsection{Quantum circuit:}   
We performed our numerical analysis with PennyLane's   ‘‘default.qubit" simulator without error mitigation.   $4$-qubit quantum states represent entities and there are $6$ ancilla qubits, $2$  of the ancilla are used for simultaneous training of $4$  triples and another $4$ qubits are used two measure  the const functions. 
We have used Adam optimiser  with   learning rate = $0.001$, loss function = mean square loss, for  epochs  $= 20$. 

\subsection{Database:} Unified Medical Language Systems(UMLS)  database with   entities = 135, relations = 46,  training triples = 5216, validation triples = 652,  and  test triples = 661  has been used in our experiment.
\subsection{Simulation result:} 
Performance of our model in fig.1 is studied  with   link prediction of the  set of  test  data  of UMLS  database.  We  take a triple from the test set and replace its tail with all the entities and find out which tail is  the most probable for that sample. The link prediction metrices  are  evaluated for all the samples in test set.  Standard matrices such as  1(Hits@1) and 10(Hits@10)  and  mean reciprocal  ratio(MRR)  are evaluated(pink rows)  and compared with the  classical state-of-the-art classical KGE models(white rows)  and our previous result(yellow rows) \cite{giri} in Table 1. While evaluating the model,  we removed any negative triples from the set of test triples generated against a test sample, which is usually called filtered data.  Note that  1(Hits@1)  and MRR  of our model  beat  NeuralLP  and  10(Hits@10)  is very close to  NeuralLP result.

\begin{table}[htbp]
 \caption{Performance of our knowledge graph embedding model  compared with  state-of-the-art  models for UMLS database.} 

\begin{center}

    \begin{tabular}{ |  p{4.0cm }  | p{0.9cm } |  p{0.9cm} | p{0.9cm} |}
    \hline
    \textbf{Models}   & \textbf{MRR}  & \textbf{Hits@1} &  \textbf{Hits@10}  \\ \hline
    
    ConvE & $95.7$  & $93.2$  &  $99.4$  \\ \hline
    
    NeuralLP  &   $77.8$  & $64.3$  &  $96.2$   \\ \hline
    
    \rowcolor{yellow} 
    Our previous model(2-qubit) & $62.0$  &  $48.9$   & $92.4$    \\  \hline
     \rowcolor{yellow} 
     Our previous model(4-qubit) & $68.1$  &  $56.3$   & $89.9$   \\  \hline
    \rowcolor{pink} 
    Our model(4-qubit, 2 layers) & $78.8$  &  $68.1$   & $95.3$   \\  \hline
    \rowcolor{pink} 
    Our model(4-qubit, 4 layers) & $79.9$  &  $70.7$   & $94.1$   \\  \hline

  \end{tabular}
 \end{center}
 \end{table} 

\section{Conclusions} \label{con}
In this article we have studied knowledge graph embedding  with variational quantum circuit.  Previous  hybrid  quantum classical models for knowledge graph embedding \cite{ma} has considered one sample at a time for the training, whereas, we have considered multiple  samples in a superposition, thereby,  further reducing  the time complexity.  Further  experiments with other KG databases and with more expressive  quantum circuits  in real quantum devices with error mitigation  will be helpful   to understand the full  potential  of our model in practical scenario.  





\end{document}